\documentclass[conference]{IEEEtran}

\usepackage[cmex10]{amsmath}
\usepackage{amssymb, amsfonts,amsthm}
\usepackage{breqn}
\usepackage{mathrsfs}
\usepackage{mdframed}
\usepackage{colonequals}

\usepackage{bm,bbm}

\usepackage{graphicx}
\usepackage{psfrag}
\usepackage{placeins}

\usepackage{enumitem}

\usepackage[square,numbers]{natbib}
\usepackage[usenames,dvipsnames]{xcolor}
\usepackage{prettyref}

\usepackage[labelfont=footnotesize, textfont=footnotesize]{caption}

\usepackage{booktabs}

\usepackage[disable]{todonotes} 

\presetkeys{todonotes}{inline,size=\small, color=green!30}{}

\newtheorem{thm}{Theorem}
\newtheorem{lem}{Lemma}

\newtheorem{cor}{Corollary}

\theoremstyle{definition}

\newtheorem{defn}{Definition}

\newtheorem{example}{Example}

\DeclareMathOperator*{\argmin}{\arg\!\min}
\DeclareMathOperator*{\argmax}{\arg\!\max}

\newcommand{\dotprod}[2]{\langle{#1},{#2} \rangle}

\newcommand{\mean}{\mathbb{E}}
\newcommand{\ExpVal}[2]{\mean{}\left[ #2 \right]}
\newcommand{\EE}[1]{\ExpVal{}{#1}}

\newcommand{\KL}{\textnormal{D}_{\scalebox{.6}{\textnormal KL}}}

\newcommand{\cfun}{\tilde{f}} 
\newcommand{\cfunat}[1]{{\cfun{}}({#1})} 

\newcommand{\textfn}[1]{{\small\textit{#1}}}

\newcommand{\calX}{\mathcal{X}}

\newcommand{\calL}{\mathcal{L}}

\newcommand{\bx}{\bm{x}}

\renewcommand{\tilde}{\widetilde}

\newcommand{\Reals}{\mathbb{R}}

\newcommand{\defined}{\triangleq}

\newcommand{\blambda}{\pmb{\lambda}}

\newcommand{\sto}{\mbox{\normalfont s.t.}}

\newcommand{\PXa}{P_{X|S=0}}
\newcommand{\PXb}{P_{X|S=1}}
\newcommand{\PYa}{P_{Y|S=0}}
\newcommand{\PYb}{P_{Y|S=1}}
\newcommand{\PXp}{\tilde{P}_{X|S=0}}
\newcommand{\PYp}{\tilde{P}_{Y|S=0}}

\newcommand{\W}{W_{Y|X}}

\newcommand{\RNum}[1]{\uppercase\expandafter{\romannumeral #1\relax}}

\begin{document}
\title{On the Direction of Discrimination: \\An Information-Theoretic Analysis of Disparate Impact in Machine Learning}
\author{%
  \IEEEauthorblockN{Hao~Wang, Berk~Ustun, and Flavio~P.~Calmon}
  \IEEEauthorblockA{John A. Paulson School of Engineering and Applied Sciences\\
                    Harvard University\\ 
                    Emails: hao\_wang@g.harvard.edu, \{berk, flavio\}@seas.harvard.edu}
}

\maketitle

\begin{abstract}

In the context of machine learning, disparate impact refers to a form of systematic discrimination whereby the output distribution of a model depends on the value of a sensitive attribute (e.g., race or gender). In this paper, we propose an information-theoretic framework to analyze the disparate impact of a binary classification model. We view the model as a fixed channel, and quantify disparate impact as the divergence in output distributions over two groups. Our aim is to find a \textit{correction function} that can perturb the input distributions of each group to align their output distributions. We present an optimization problem that can be solved to obtain a correction function that will make the output distributions statistically indistinguishable. We derive closed-form expressions to efficiently compute the correction function, and demonstrate the benefits of our framework on a recidivism prediction problem based on the ProPublica COMPAS dataset.

\end{abstract}

\section{Introduction}
\label{Sec::Introduction}

Machine learning (ML) models aim to exploit biases in the training data to predict an outcome of interest.  In many real-world applications, however, effective prediction should not be achieved by discriminating on a \textit{sensitive attribute}, such as race or gender \cite{romei2014multidisciplinary,hajian2016algorithmic}.

Discrimination can occur directly when a sensitive attribute is used as an input to the model, known as \emph{disparate treatment}. More pervasive today is a phenomenon known as \emph{disparate impact} \cite{barocas2016disparate}, where a sensitive attribute is omitted from the model, but still affects its predictions through correlations with ``proxy" variables (e.g., income, education level). The potential to discriminate by proxy is not unique to ML. In the United States, for example, racial minorities were indirectly denied financial services by exploiting correlations between race, address, and income -- a practice known as \textit{redlining} \cite{hunt2005redlining}. 

Disparate impact can arise as an artifact of empirical loss minimization when a sensitive attribute is valuable for prediction and can be approximately inferred using other \textit{proxy} variables in the training data.
A large body of recent work has documented this phenomenon in real-world applications ranging from online advertising \cite{sweeney2013discrimination} to recidivism prediction \cite{angwin2016machine}. When used in human or algorithmic decision-making, models with disparate impact may violate anti-discrimination laws \citep{barocas2016disparate} and inadvertently amplify societal biases \cite{ensign2017decision}.

These issues have motivated a growing stream of technical work on disparate impact in ML, focusing on topics such as: (i) how to identify and quantify disparate impact \cite{feldman2015certifying,adler2016auditing,adebayo2016iterative,simoiu2017problem}; (ii) how to train models that mitigate disparate impact \citep{kamiran2009classifying,zemel2013learning,calmon2017optimized}; and (iii) how to identify causal factors of discrimination \cite{kilbertus2017avoiding}. The present work is inscribed within the first research direction. 

In this paper, we consider the disparate impact problem from an information-theoretic perspective. Our goal is to derive a \textit{correction function} that can be used to identify features that act as proxies of a sensitive attribute for a fixed prediction model. We first present a correction function that has an information-theoretic interpretation in terms of error exponents of binary hypothesis testing (Section \ref{Sec::Framework}). We then derive closed-form expressions for the correction function that can easily be computed using the prediction model, which uses the features $X$ to predict the outcome $Y$, and a group membership distribution, which uses the features $X$ to ``predict" the sensitive attribute $S$ (Section \ref{Sec::Results}).
Our approach is inspired by recent work in information-theoretic privacy \cite{liao2016hypothesis}, which takes a similar route to analyzing the behavior of error exponents under small perturbations.
We illustrate our framework on a recidivism prediction problem derived from the ProPublica COMPAS dataset \cite{angwin2016machine} (Section \ref{Sec::Experiments}). 

\section{Framework}
\label{Sec::Framework}

We consider a channel $\W$, which takes as input a vector of $d$ random variables $X = (X_1,\ldots,X_d) \in \mathcal{X}$ and produces as output a random variable $Y \in \mathcal{Y}$. We assume that the support sets 
$\mathcal{X}$ and $\mathcal{Y}$ are finite. In practice, $\W$ represents a predictive model (e.g., a linear classifier to predict recidivism), $X$ represents a vector of features (e.g., \textfn{Age}, \textfn{Salary}), $Y$ represents the predicted output of $\W$ given $X$ (e.g., $Y = 1$ iff the model predicts that a prisoner with features $X$ will commit a crime after being released from prison). 

We seek to characterize differences in the output distribution of the channel $\W$ with respect to a \emph{sensitive attribute} $S$.  We focus on the case where the sensitive attribute is binary $S\in \{0,1\}$, and use $\PXa, \PXb$ and $\PYa, \PYb$ to denote the conditional distributions of inputs and outputs, respectively. A channel   $\W$ is said to have \emph{disparate impact} with respect to $S$ when $\PYa \neq \PYb$.
We assume that $\W$ does not use the sensitive attribute $S$, as doing so would violate legal constraints in applications such as hiring and credit scoring (see, e.g., \cite{barocas2016disparate}). In this setting, the Markov condition $S\to X\to Y$ ensures that $\PYa=\W\circ\PXa$ and $\PYb=\W\circ\PXb$. Thus, disparate impact  occurs only when $\PXa \neq \PXb$.

\begin{figure}[!tb]
  \centering
  \includegraphics[width=0.48\textwidth]{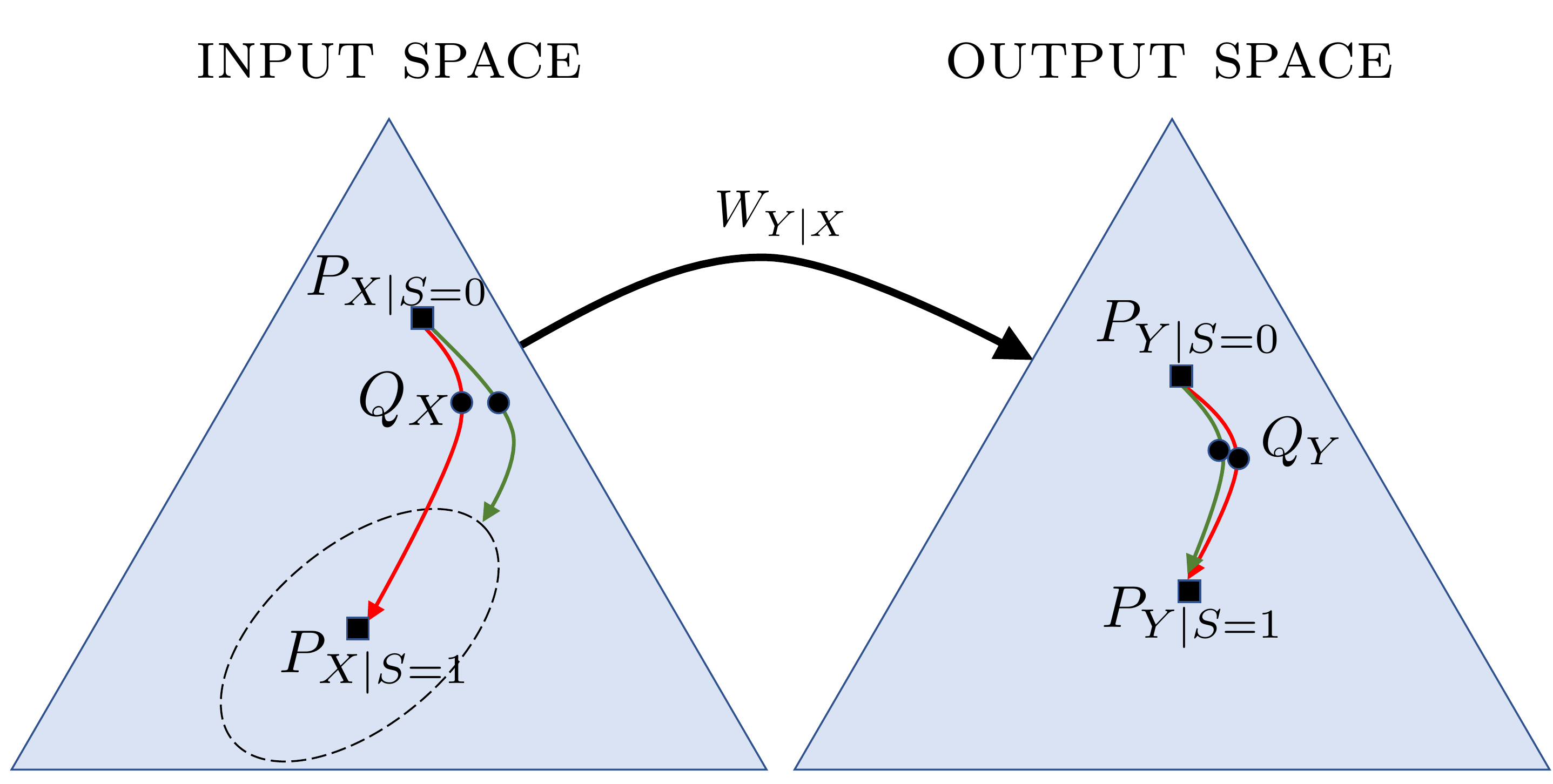}
  \caption{Two correction paths on the probability simplex. The green line depicts a path that reduces disparate impact by aligning only output distributions. The red line depicts a path that reduces disparate impact by aligning the input distributions and output distributions.}
  \label{Fig::CorrectionPaths}
\end{figure}

Given a channel $\W$, disparate impact can be reduced by perturbing $\PXa$ to a new distribution $Q_X$ so that the resulting output distribution $Q_Y=\W\circ Q_X$ is ``closer'' to $\PYb$ (cf. Fig.  \ref{Fig::CorrectionPaths}). Intuitively, larger disparities between output distributions require larger perturbations, and the direction between $Q_X$ and $\PXa$ reflects which components of $X$ contribute to this disparity. In what follows, we will define this setup formally.

\begin{defn}
Given a convex divergence metric $J(\cdot\|\cdot)$ (e.g., total variation or KL-divergence),  a fixed joint distribution $P_{S,X,Y}$ such that $S\rightarrow X\rightarrow Y$, and $\blambda\defined (\lambda_1,\lambda_2,\lambda_3,\lambda_4)$ such that $\lambda_i\geq 0$, the objective function is defined as:
\begin{align}
\label{eq:obj}
\begin{split}
    L_\blambda(Q_X) \defined & \phantom{+} \lambda_1 J(Q_X\|\PXa)+\lambda_2 J(Q_X\|\PXb) \\
     &+ \lambda_3 J(Q_Y\|\PYa)+\lambda_4 J(Q_Y\|\PYb).
\end{split}
\end{align}

The \textit{correction path} between $\PXa$ and $\PXb$ is defined as an optimal solution to \eqref{eq:obj} for a fixed value of $\blambda$:
\begin{align}
L_{\blambda}^{\text{opt}}\defined
\min_{Q_X}~& L_\blambda(Q_X).
    \label{eq:corrPath}
\end{align}
\end{defn}
From standard convexity results \citep[e.g.,][]{cover2006elements}, as the values of $\lambda_i$ are changed, the distribution $Q_X$ that minimizes \eqref{eq:corrPath} describes the lower boundary of the set %
\begin{align*}
    \{(& J(Q_X\|\PXa),J(Q_X\|\PXb),\\
    &J(Q_Y\|\PYa),J(Q_Y\|\PYb) ) \mid \text{supp}(Q_X)\subseteq \mathcal{X}\}.
\end{align*} 
Solving \eqref{eq:corrPath} will produce different paths on the probability simplex depending on the components of $\blambda$ (see  Fig.~\ref{Fig::CorrectionPaths}). When $\lambda_2=\lambda_4=0$, the optimal solution is $Q_X=\PXa$. When $\lambda_2=\lambda_3=0$, $Q_X$ will traverse the shortest path (as measured by $J(\cdot\|\cdot)$) between $\PXa$ and the set $\{P_X\mid J(\W\circ P_X\|\PYb)=0\}$ (i.e., the green line in Fig.~\ref{Fig::CorrectionPaths}). Note that this path transforms $\PXa$ into a distribution devoid of disparate impact, but potentially different from $\PXb$. Perhaps of greater interest is the path obtained when $\lambda_3=0$ (i.e., the red line in Fig.~\ref{Fig::CorrectionPaths}). In this case, varying $\lambda_1,\lambda_2, \lambda_4 \geq 0$ corresponds to traversing between $\PXa$ and $\PXb$, while controlling the similarity of the induced distribution on $Y$.

Our goal is to produce a \textit{correction function} that indicates which components of $X$ contribute to disparate impact. More precisely, the correction function is a local (multiplicative) perturbation of $\PXa$ that decreases the objective function \eqref{eq:obj} the most (see Definition~\ref{Def:perturb}). This definition leads to correction functions that can be cast in terms of predictive models for $S$ and $Y$ given $X$, as we show in Section~\ref{Sec::Results}.
\begin{defn}
\label{Def:perturb}
For a given function $f\in \calL(\PXa)$, we define the perturbed distribution $\PXp$ as
\begin{align}
\label{Eq::Perturb_dist}
\PXp(\bx)\defined \PXa(\bx)(1+\epsilon f(\bx)),
\end{align}
where $\epsilon>0$ is chosen so that $\PXp$ is a valid probability distribution, and
\begin{equation*}
\begin{aligned}
&\calL(\PXa)\defined \\
&\left\{f:\calX\to \Reals \mid  \EE{f(X)|S=0}=0, \EE{f(X)^2|S=0} =1 \right\}.
\end{aligned}
\end{equation*}
Moreover, $\PYp \defined \W\circ\PXp$.
\end{defn}
Next, we denote $\Delta_{\blambda}(f)$ as the decrease in the objective function \eqref{eq:obj} by locally perturbing the distribution $\PXa$.
\begin{defn}
\label{Def:lambda}
For a given $\blambda$, we define $\Delta_{\blambda}(f)$ as
\begin{equation}
\label{Eq:Delta_defn}
\Delta_{\blambda}(f) \defined \lim_{\epsilon \to 0^+} \frac{L_{\blambda}(\PXp) - L_{\blambda}(\PXa)}{\epsilon}.
\end{equation}
\end{defn}
\begin{defn}
For a given $\blambda$, the correction function $\cfun$ is the minimizer of  $\Delta_{\blambda}(f)$:
\begin{equation}
\label{eq::defn_correctionfunction}
    \cfun = \argmin_{f\in \calL(\PXa)}\Delta_{\blambda}(f).
\end{equation}
\end{defn}
We remark that the influence of local perturbations on probability distributions has been studied both in statistics (see, e.g., \cite{huber2011robust}) and information theory (see e.g., \cite{borade2008euclidean}).
\subsection*{Connections to Binary Hypothesis Testing}
In the remainder of this paper, we consider settings where the outcome variable is binary $Y \in \{0,1\}$, and the divergence measure is the KL-divergence $J(\cdot\|\cdot)=\KL(\cdot\|\cdot)$. In this case, the objective function~\eqref{eq:obj} can be expressed as:
\begin{align}
\begin{split}
L_{\blambda}(Q_X) &= \lambda_1 \KL(Q_X\|\PXa)+\lambda_2 \KL(Q_X\|\PXb) \\ 
 & + \lambda_3 \KL(Q_Y\|\PYa)+ \lambda_4\KL(Q_Y\|\PYb). \label{eq:obj_KL}
\end{split}
\end{align}

Our choice of $\KL(\cdot\|\cdot)$ is motivated by its relationship with the error exponent in hypothesis testing (see Ch 11 in \cite[][]{cover2006elements}, \cite{blahut1974hypothesis,tuncel2005error}). Specifically, when $\lambda_3=\lambda_4=0$ in \eqref{eq:corrPath}, the correction path describes the best trade-off (in terms of the first order term in the exponent) between the Type I and Type II error for a hypothesis test that seeks to distinguish $S$ given an observation of $X$. The optimal value of \eqref{eq:corrPath} can then be expressed in terms of the R\'enyi's $\alpha$-divergence (cf. \cite[Section II-A]{tuncel2005error}), i.e., $D_\alpha (\PXa\|\PXb)$. 
In other words, under the choice of KL-divergence the correction path can be understood as the trade-off between error exponents of two (independent) binary hypothesis tests: one to distinguish $S$ from $X$; and another to distinguish $S$ from $Y$.

\section{Main Results}
\label{Sec::Results}
In this section, we derive closed-form expressions for the correction function for the objective function defined in \eqref{eq:obj_KL}. We leverage the connection between small perturbations of KL-divergences and maximal correlation \citep[as noted, for example, by][]{gohari2012evaluation,anantharam2013maximal}.

Our main results consist of Theorems \ref{Thm::AlgorithmMin} and \ref{thm::SimplifyBasis}, where we prove that the correction function that minimizes \eqref{eq::defn_correctionfunction} is a linear combination of two components: $f_l$, which aligns the perturbed input distribution with $\PXb$; and  $f_m$, which aligns the corresponding output distribution with $\PYb$. In Theorem \ref{thm::SimplifyBasis}, we show that $f_l$ and $f_m$ (and thus $\cfun{}$) can be expressed in terms of the group membership distribution $P_{S|X}$ and the channel $\W$. This result has an important practical benefit: it allows us to compute the correction function $\cfun{}$ directly using only $P_{S|X}$ and  $\W$, without computing the complete joint distribution $P_{S,X,Y}$. In what follows, we present a formal statement of these results.  

We start our derivation of the correction function by providing a simplified expression for  $\Delta_{\blambda}(f)$ in Lemma \ref{Lemma::delta}.

\begin{lem}
\label{Lemma::delta}
For a given $\blambda$, $\Delta_{\blambda}(f)$ can be expressed as 
\begin{align}
\label{Eq::lem_delta}
\begin{split}
\Delta_{\blambda}(f) 
= &\,\lambda_2 \EE{f(X)\log\frac{\PXa(X)}{\PXb(X)}\Bigg|S=0}\\ 
  &\,+\lambda_4 \EE{g(Y)\log\frac{\PYa(Y)}{\PYb(Y)}\Bigg|S=0},
\end{split}
\end{align}
where $g(y)=\EE{f(X)|Y=y,S=0}$.
\end{lem}

%

Next, we introduce definitions used to derive the correction function.
\begin{defn}
The \textit{log-likelihood ratio functions} $f_l$ and $g_l$ are given by
\begin{align}
f_l(\bx) &\defined \log\frac{\PXa(\bx)}{\PXb(\bx)} - \EE{\log\frac{\PXa(X)}{\PXb(X)}\Bigg|S=0},\\
g_l(y) &\defined \log\frac{\PYa(y)}{\PYb(y)} - \EE{\log\frac{\PYa(Y)}{\PYb(Y)}\Bigg|S=0}.
\end{align}
\end{defn}

\begin{defn}
\label{defn::maxmalcorrelation}
The \textit{maximal correlation} (see, e.g., \cite{renyi1959measures}) between $X$ and $Y$ given $S=0$ is defined as 
\begin{align*}
\rho_m(\PXa;\W) \defined \max_{\substack{f\in \calL(\PXa)\\ g\in \calL(\PYa)}} \EE{f(X)g(Y)|S=0}.
\end{align*}
The maximal correlation can be equivalently given by
\begin{equation*}
\begin{aligned}
    \rho_m(\PXa;\W)=\sqrt{\EE{\EE{g_m(Y)|X,S=0}^2\Big|S=0}}.
\end{aligned}
\end{equation*}
\end{defn}
We refer to the functions that attain the maximum as the \textit{principal functions} and denote them as $(f_m, g_m)$.

Theorems \ref{Thm::AlgorithmMin} and \ref{thm::SimplifyBasis} characterize the correction function $\cfun$. Theorem~\ref{Thm::AlgorithmMin} shows that the correction function is the linear combination of the log-likelihood ratio function $f_l$ and the principal function $f_m$. Theorem~\ref{thm::SimplifyBasis} shows that $f_l$ and $f_m$ can be expressed in terms of the group membership distribution $P_{S|X}$ and the channel $\W$, respectively.
\begin{thm}
\label{Thm::AlgorithmMin}
Given $\blambda$, the correction function $\cfun$ has the form
\begin{align}
\label{Eq::CorrectionFunction_Component}
\cfun =  n_l f_l + n_m f_m,
\end{align}
where $n_l$ and $n_m$  are constants computed as:
\begin{align*}
n_l& = \frac{-\lambda_2}{\sqrt{\left(\lambda_2  a_1+\lambda_4\rho_m(\PXa;\W)b_1\right)^2+(\lambda_2 a_2)^2}}, \\
n_m& = \frac{-\lambda_4\rho_m(\PXa;\W)b_1}{\sqrt{\left(\lambda_2 a_1+\lambda_4\rho_m(\PXa;\W)b_1\right)^2+(\lambda_2 a_2)^2}},\\
a_1&=\EE{f_l(X)f_m(X)|S=0},\\
a_2&=\sqrt{\EE{\left(f_l(X)-a_1f_m(X)\right)^2\Big|S=0}},\nonumber\\ 
b_1&=\EE{g_l(Y)g_m(Y)|S=0}.
\end{align*}
\end{thm}
%
%
\begin{thm}
\label{thm::SimplifyBasis}
The log-likelihood ratio function $f_l$ and the principal function $f_m$ can be expressed as
\begin{align}
&f_l(\bx) = \log\frac{P_{S|X}(0|\bx)}{P_{S|X}(1|\bx)} - \EE{\log\frac{P_{S|X}(0|X)}{P_{S|X}(1|X)}\Bigg|S=0},\\
&f_m(\bx)= \frac{(g_m(1)-g_m(0))\W(1|\bx) +g_m(0)}{\rho_m(\PXa;\W)}. \label{eq:f_mKL}
\end{align}
Here, $g_m(0)=\sqrt{p/(1-p)}$ and $g_m(1)=-\sqrt{(1-p)/p}$ where $p\defined\PYa(1)$.
\end{thm}
%
%
Combining Theorems \ref{Thm::AlgorithmMin} and \ref{thm::SimplifyBasis}, we obtain a closed-form expression for the correction function $\cfun$.
Note that, due to our definition of $\cfun$ in terms of local perturbations, the correction function does not depend on $\lambda_1$ and $\lambda_3$ as in \eqref{eq:obj_KL}. 
In Corollary~\ref{Cor::ObjectiveValue}, we provide an expression for $\Delta_{\blambda}(\cfun)$ in this case:
\begin{cor}
\label{Cor::ObjectiveValue}
For a given $\blambda$,  $\Delta_{\blambda}(\cfun)$ is given by
\begin{equation*}
\begin{aligned}
\Delta_{\blambda}(\cfun)
= -\sqrt{\left(\lambda_2 a_1+\lambda_4\rho_m(\PXa;\W)b_1\right)^2+(\lambda_2 a_2)^2}
\end{aligned}
\end{equation*}
where $a_1$, $a_2$, $b_1$ are defined in Theorem~\ref{Thm::AlgorithmMin}.
\end{cor}

We now instantiate our results for the case where $\lambda_2=0$, where our objective is to align only the output distributions (i.e., the green line in Fig. \ref{Fig::CorrectionPaths}). As expected, the following corollary shows that the correction function $\cfun$ under this scenario is (up to a sign difference) the principal function $f_m$.
\begin{cor}
\label{Cor::RxequalQx}
When $\lambda_2=0$, the correction function $\cfun$ is:
\begin{align}
\cfun = -\textnormal{sign}\left(\EE{g_l(Y)g_m(Y)|S=0}\right)f_m
\end{align}
and
\begin{align*}
&\Delta_{\blambda}(\cfun) \\
&= - \lambda_4 \rho_m(\PXa;\W) \sqrt{\textnormal{Var}\left[\log\frac{\PYa(Y)}{\PYb(Y)}\Bigg|S=0\right]}.
\end{align*}
\end{cor}
We conclude this section with Example \ref{ex::Logistic2}, where we compute $\cfun$ when $P_{S|X}$ follows a logistic distribution.
\begin{example}
\label{ex::Logistic2}
When $P_{S|X}(1|\bx)=(1+\exp(\theta_0+\dotprod{\bm{\theta}}{\bx})^{-1}$, the log-likelihood ratio function $f_l$ is the linear function
\begin{equation}
\label{eq:fl_log}
f_l(\bx)= \dotprod{\bm{\theta}}{\bx} - \sum_{i=1}^d \EE{\theta_i X_i\Big|S=0}.
\end{equation}
Combining \eqref{Eq::CorrectionFunction_Component}, \eqref{eq:f_mKL} and \eqref{eq:fl_log}, $\cfun$ can be expressed as a linear combination of $\W$, a linear function of $\bx$, and a constant term. Further, when $P_{S|Y}(1|y)=(1+\exp(\gamma_0+\gamma_1y))^{-1}$,  $\Delta_{\blambda}(f)$ can be expressed as
\begin{align*}
\label{Eq::PRQh_logistic}
\Delta_{\blambda}(f)
=\lambda_2 \sum_{i=1}^d \theta_i \EE{f(X)X_i|S=0} + \lambda_4\gamma_1\EE{g(Y)Y|S=0}.
\end{align*}
for any perturbation $f$, including  $\cfun$.
\end{example}

\section{Numerical Experiments}
\label{Sec::Experiments}
We now discuss a numerical experiment where we compute correction functions for a recidivism prediction model. We consider the ProPublica COMPAS dataset \cite{angwin2016machine}, which contains information on the criminal history and demographic makeup of prisoners in Brower County, Florida from 2013--2014. Our goal is to illustrate the technical feasibility of our approach on a real-world dataset, and to show that the correction function can be computed using standard predictive models for $S$ and $Y$ given $X$ (i.e., without the need to compute the  distribution $P_{S,X,Y}$). We provide code to reproduce our analysis at \cite{github2018}.

\subsection*{Setup}

We restrict our analysis to individuals who are African American ($S = 0$) or Caucasian ($S = 1$). We process the raw dataset by dropping records with missing information and converting categorical variables to numerical values. Our final dataset contains 5278 records (3175 African American + 2103 Caucasian), where the record for individual $i$ consists of a feature vector $\bm{x}_i = (\textfn{Age}, \textfn{ChargeDegree}, \textfn{Sex}, \textfn{PriorCounts}, \textfn{LengthOfStay})$, and an outcome variable, set as $y_i = 1$ iff they are arrested for a crime within 2 years of release from prison.

We use the entire dataset to train two logistic regression models: (i) $\W$, which uses the features to predict the outcome; (ii) $P_{S|X}$, which uses the features to predict group membership. Although $\W$ does not use $S$ as an input, it has significant disparate impact over $S$, assigning higher scores on average to African Americans compared to Caucasians ($\mathbb{E}[Y|S=0] = 0.543$ vs. $\mathbb{E}[Y|S=1] = 0.438$). Using $\W$ and $P_{S|X}$, we apply Theorems \ref{Thm::AlgorithmMin} and \ref{thm::SimplifyBasis} to compute the correction function $\cfun$ for $\lambda_2=\lambda_4=1$.

\subsection*{Results}

\begin{figure}[!b]
  \centering
  \includegraphics[width=0.35\textwidth]{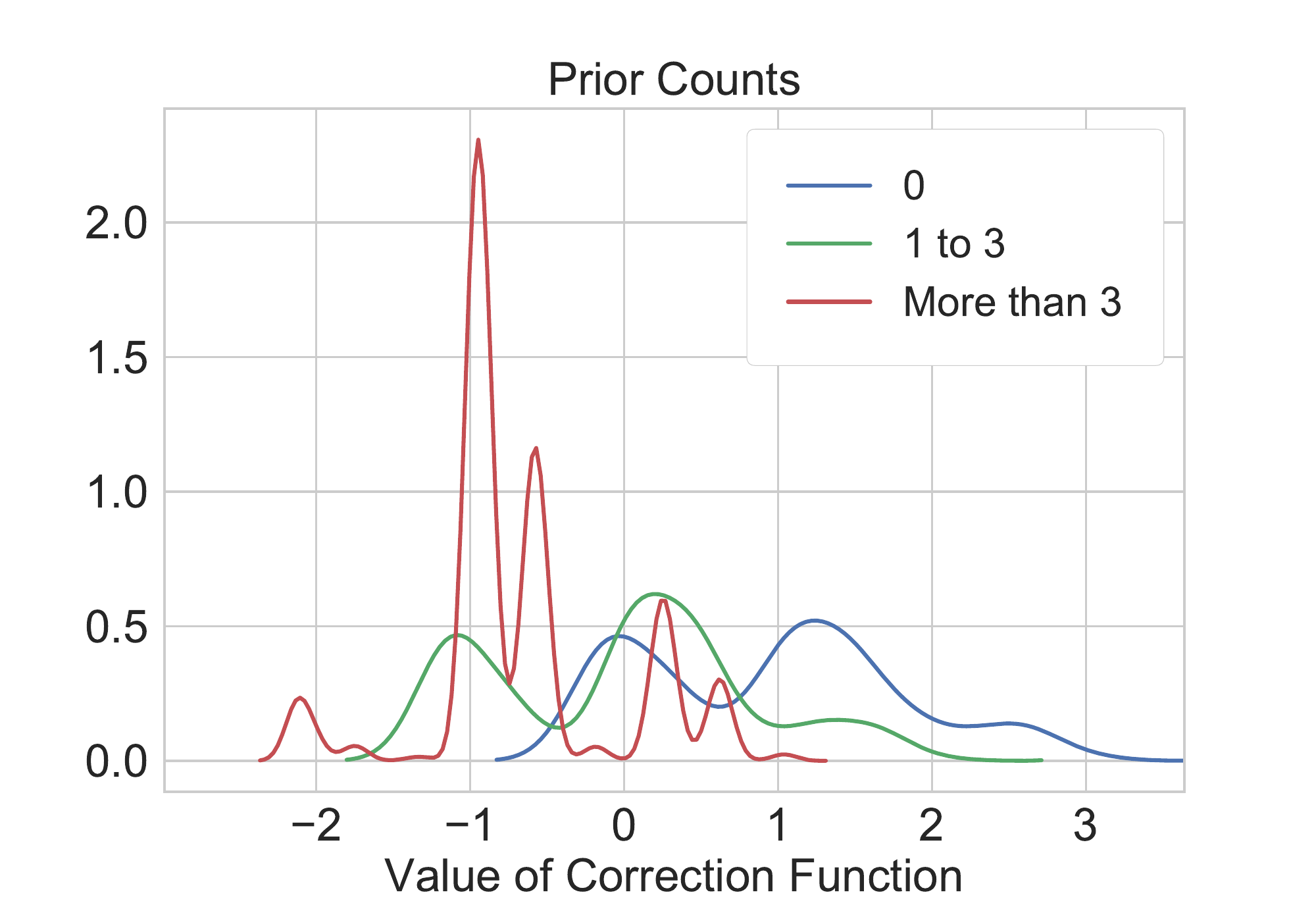}
  \includegraphics[width=0.325\textwidth]{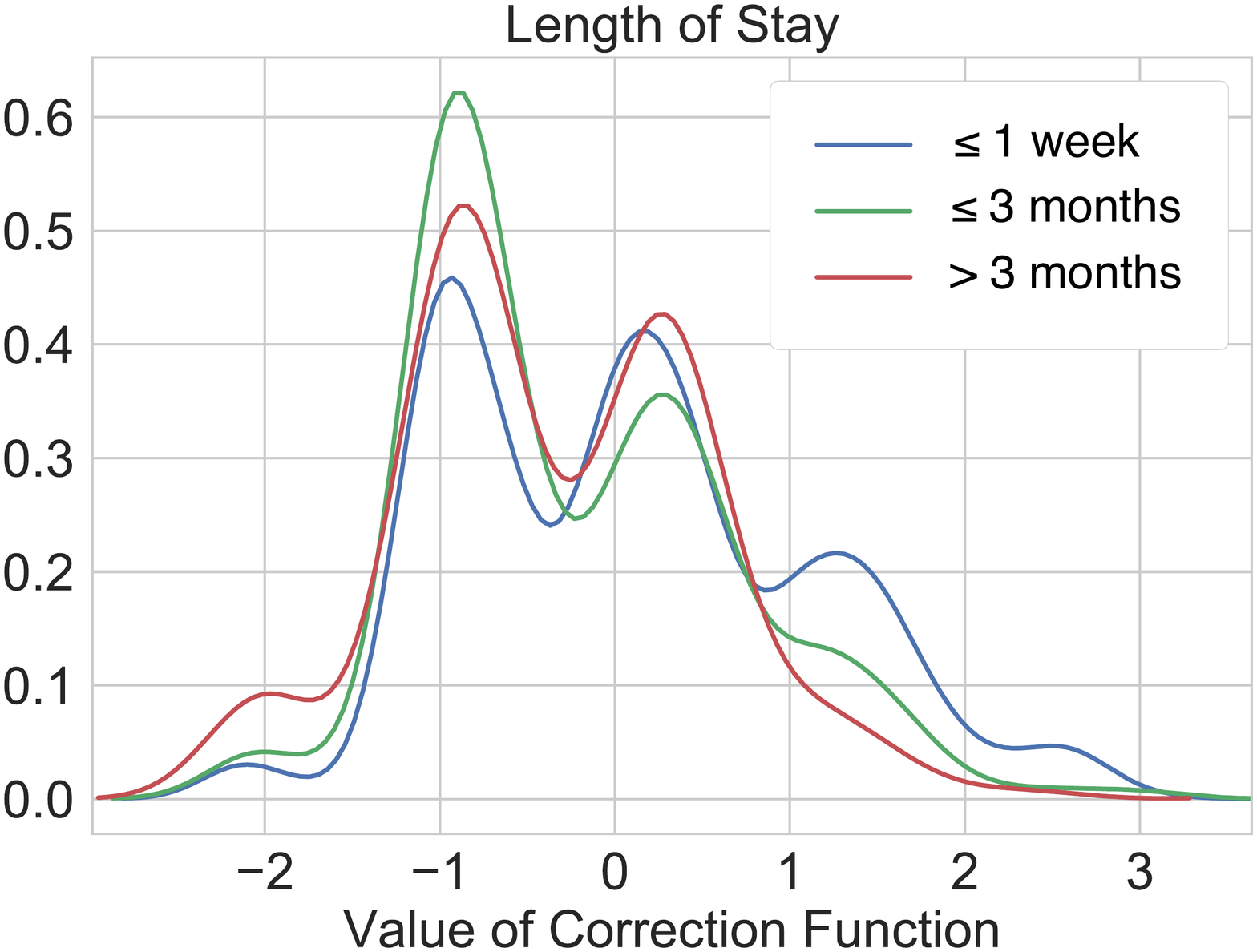}
  \caption{Empirical  distributions of the correction function $\cfun$ conditioned on distinct values of \textit{PriorCounts} (top) and \textit{LengthOfStay} (bottom). We use a kernel density estimator to smooth the histograms.} 
  \label{Fig::DistributionPlots_COMPAS}
\end{figure}

In Fig.~\ref{Fig::DistributionPlots_COMPAS}, we show how the correction function can identify features that contribute to the disparate impact of a predictive model. Here, we plot the conditional distribution of $\cfun$ for distinct values of $\textfn{PriorCounts}$ and $\textfn{LengthOfStay}$. As shown, the distribution of $\cfun{}$ is similar across all values of $\textfn{LengthOfStay}$, which suggests that $\textfn{LengthOfStay}$ does not affect the disparate impact of the model. In contrast, the distribution of $\cfun{}$ differs based on the value of \textfn{PriorCounts} (see e.g., the differences between $\textfn{PriorCounts} > 3$). This suggests that the model may be using \textfn{PriorCounts} to discriminate between African Americans and Caucasians.

In Table \ref{Table::Prototypes_COMPAS}, we show prototypical examples for the correction function, which correspond to feature vectors for which the correction function $\cfun{}$ attains its maximum value/minimum value/value closest to 0. Recalling that the $\cfun{}$ represents a local perturbation that minimizes the objective in \eqref{Eq:Delta_defn}, we see that the disparity in the output distributions of $\W$ is maximal for African American males that are under 25 years old, with $> 3$ priors, and charged with a felony (middle column). This is to be expected, as the training dataset for $\W$ shows a large correlation between \textfn{PriorCounts} and the outcome, and higher \textfn{PriorCounts} for African Americans on average. 
\begin{table}[!tb]
\centering
{\scriptsize
\begin{tabular}{lrrr}
\toprule
\textsc{Features} & %
$\argmax{\cfunat{\bm{x}}}$ &%
$\argmin{\cfunat{\bm{x}}}$ &%
$\argmin{|\cfunat{\bm{x}}|}$ \\
\toprule
\textit{Age}   & $>45$              & $<25$          & 25 to 45 \\
\midrule
\textit{ChargeDegree}  & Misdemeanor        & Felony         & Felony \\
\midrule
\textit{Sex}            & Female             & Male           & Male \\
\midrule
\textit{PriorCounts}   & $0$                & $>3$           & 1 to 3 \\
\midrule
\textit{LengthOfStay} & $<3$ Months  & $<$ Week & $<$ Week \\
\bottomrule
\end{tabular}
}
\caption{Prototypical examples for the correction function $\cfun$. We show the features of African American entries that attain the maximum, minimum, and smallest values of $\cfun$.}
\label{Table::Prototypes_COMPAS}
\end{table}  

\clearpage
\section{Discussion}
\label{Sec::Discussion}

Disparate impact in machine learning is a critical issue with important societal implications. In this paper, we proposed an information-theoretic framework to study disparate impact. We derived a correction function that reflects how components of the input variables $X$ affect the disparity in the output distributions. We then demonstrated how our framework could be used on a recidivism prediction application derived from a real-world dataset. Interesting directions for future work include extending our analysis to a broader class of predictive models, and using correction functions to design machine learning algorithms that mitigate disparate impact. We are confident that information-theoretic tools can inspire exciting new  solutions to the problem.

\section*{Acknowledgments}
F.P. Calmon would like to thank the Harvard Dean's Competitive Fund for Promising Scholarship for supporting this research.

\bibliographystyle{IEEEtran}
{\small
\bibliography{reference}
}

\clearpage

\appendices
\section{Proofs}
\label{Appendix::Proofs}

\subsection{Proof of Lemma \ref{Lemma::delta}}
\label{Appendix::ProofLemmaObjectiveFunction}
\begin{proof}
First, note that we can compute the distribution $\PYp$ by passing $\PXp$ through the given channel $\W$ and get the following expression.
\begin{align*}
\PYp(y) 
=& \sum_{\bx} \W(y|\bx)\PXa(\bx)(1+\epsilon f(\bx))\\
=& \PYa(y)+\epsilon \sum_{\bx} \W(y|\bx)\PXa(\bx)f(\bx)\\
=& \PYa(y)+\epsilon \sum_{\bx} P_{X|Y,S=0}(\bx|y)\PYa(y)f(\bx)\\
=& \PYa(y)\left(1+\epsilon \EE{f(X)|Y=y,S=0}\right)\\
=& \PYa(y)(1+\epsilon g(y)).
\end{align*}

By the definition of $\PXp$, we can compute the KL-divergence between $\PXp$ and $\PXb$ in the following way.
\begin{equation}
\label{Eq::KL_1}
\begin{aligned}
&\KL(\PXp\|\PXb)\\
&=\sum_{\bx} \PXa(\bx)(1+\epsilon f(\bx))\log\frac{\PXa(\bx)(1+\epsilon f(\bx))}{\PXb(\bx)}\\
&=\KL(\PXa\|\PXb)\\
&\quad+\epsilon \sum_{\bx} \PXa(\bx)f(\bx)\log\frac{\PXa(\bx)}{\PXb(\bx)}\\
&\quad+\sum_\bx \PXa(\bx)(1+\epsilon f(\bx))\log(1+\epsilon f(\bx))\\
&= \KL(\PXa\|\PXb)\\
&\quad+\epsilon\EE{f(X)\log\frac{\PXa(X)}{\PXb(X)}\Bigg|S=0}\\
&\quad+\epsilon\sum_\bx \PXa(\bx)(1+\epsilon f(\bx))f(\bx) + O(\epsilon^2)\\
&= \KL(\PXa\|\PXb)\\
&\quad+\epsilon\EE{f(X)\log\frac{\PXa(X)}{\PXb(X)}\Bigg|S=0}\\
&\quad+\epsilon\EE{f(X)|S=0}+\epsilon^2\EE{f(X)^2|S=0}+ O(\epsilon^2)\\
&= \KL(\PXa\|\PXb)\\
&\quad+\epsilon\EE{f(X)\log\frac{\PXa(X)}{\PXb(X)}\Bigg|S=0}+ O(\epsilon^2).
\end{aligned}
\end{equation}
Following similar computation, we have
\begin{equation}
\label{Eq::KL_2}
    \KL(\PXp\|\PXa) =  O(\epsilon^2).
\end{equation}
Since $\EE{g(Y)|S=0}=0$, we have
\begin{align}
\label{Eq::KL_3}
\begin{split}
&\KL(\PYp\|\PYb)-\KL(\PYa\|\PYb)\\
&=\epsilon\EE{g(Y)\log\frac{\PYa(Y)}{\PYb(Y)}\Bigg|S=0}+O(\epsilon^2).
\end{split}
\end{align}
Also,
\begin{align}
\label{Eq::KL_4}
\begin{split}
\KL(\PYp\|\PYa)=O(\epsilon^2).
\end{split}
\end{align}
Combining \eqref{Eq::KL_1}, \eqref{Eq::KL_2}, \eqref{Eq::KL_3}, \eqref{Eq::KL_4} together and letting $\epsilon\to 0$, we obtain
\begin{align*}
\Delta_{\blambda}(f) 
= &\lambda_2\EE{f(X)\log\frac{\PXa(X)}{\PXb(X)}\Bigg|S=0}\\
& + \lambda_4\EE{g(Y)\log\frac{\PYa(Y)}{\PYb(Y)}\Bigg|S=0}.
\end{align*}
\end{proof}

\subsection{Proof of Theorem \ref{Thm::AlgorithmMin}}
\label{Appendix::ProofThmMinimizer}

\begin{proof}
Note that when $Y$ is a binary random variable, for any function $g(y)$ with $\EE{g(Y)}=0$, we have that $g(Y)=\EE{g(Y)g_m(Y)}g_m(Y)$. Furthermore, for any function $f(\bx)$ with $\EE{f(X)}=0$, if $\EE{f(X)f_m(X)}=0$, then $\EE{f(X)|Y}=0$.

We define
\begin{align}
f_L(\bx) \defined \frac{f_l(\bx)-a_1f_m(\bx)}{a_2},
\end{align}
when $a_2 \neq 0$. When $a_2=0$, we can choose an arbitrary function $f_L$ such that $f_L\in \calL(\PXa)$ and $\EE{f_L(X)f_m(X)|S=0}=0$. Note that $f_L\in \calL(\PXa)$ from the definition. Furthermore, we have $g_l(y)=b_1g_m(y)$ with $b_1=\EE{g_l(Y)g_m(Y)|S=0}$.
Similarly, 
\begin{align}
f(\bx) &= m_1f_m(\bx)+m_2f_L(\bx)+m_3f_r(\bx) \\ 
\intertext{where,}
m_1 & = \EE{f(X)f_m(X)|S=0}, \nonumber \\
m_2 & = \EE{f(X)f_L(X)|S=0}, \nonumber \\
m_3 & = \sqrt{\EE{\left(f(X)-m_1f_m(X)-m_2f_L(X)\right)^2\Big|S=0}}. \nonumber
\end{align}
\begin{align*}
f_r(\bx) \defined \frac{f(\bx)-m_1f_m(\bx)-m_2f_L(\bx)}{m_3},
\end{align*}
when $m_3\neq 0$. When $m_3=0$, we can choose an arbitrary function $f_r$ such that $f_r\in \calL(\PXa)$ and $\EE{f_r(X)f_m(X)|S=0}=\EE{f_r(X)f_L(X)|S=0}=0$. Note that $f_r\in \calL(\PXa)$ following the definition. Since, by the definition of $f_L$ and $f_r$, $\EE{f_m(X)f_L(X)|S=0}=0$ and $\EE{f_m(X)f_r(X)|S=0}=0$, then
\begin{align*}
g(y)&=\EE{f(X)|Y=y,S=0}\\
&=\rho_m(\PXa;\W)m_1g_m(y).
\end{align*}
Therefore, following previous discussions and using Lemma \ref{Lemma::delta}, we have
\begin{align*}
&\Delta_{\blambda}(f)\\
&=  \lambda_2\EE{f(X)\log\frac{\PXa(X)}{\PXb(X)}\Bigg|S=0}\\
& + \lambda_4\EE{g(Y)\log\frac{\PYa(Y)}{\PYb(Y)}\Bigg|S=0} \\
&= \lambda_2\EE{f(X)f_l(X)\Bigg|S=0} + \lambda_4\EE{g(Y)g_l(Y)\Bigg|S=0}\\
&= \lambda_2 a_1m_1+\lambda_2 a_2m_2+\lambda_4\rho_m(\PXa;\W)b_1m_1 .
\end{align*}
Since $f\in \calL(\PXa)$, we have that $m_1^2+m_2^2+m_3^2=1.$ Accordingly, we can minimize $\Delta_{\lambda}(f)$ by solving the optimization problem:
\begin{align*}
\min_{m_1,m_2} &\qquad \left(\lambda_2 a_1+\lambda_4\rho_m(\PXa;\W)b_1\right)m_1+\lambda_2 a_2m_2\\
\sto & \qquad m_1^2+m_2^2+m_3^2=1.
\end{align*}
By the Cauchy-Schwarz inequality, the minimal value is $-\sqrt{\left(\lambda_2 a_1+ \lambda_4\rho_m(\PXa;\W)b_1\right)^2+(\lambda_2 a_2)^2}$ which is achieved by setting
\begin{align*}
m_1 &= \frac{-\left(\lambda_2 a_1+\lambda_4\rho_m(\PXa;\W)b_1\right)}{\sqrt{\left(\lambda_2 a_1+\lambda_4\rho_m(\PXa;\W)b_1\right)^2+(\lambda_2 a_2)^2}},\\
m_2 &= \frac{-\lambda_2 a_2}{\sqrt{\left(\lambda_2 a_1+\lambda_4\rho_m(\PXa;\W)b_1\right)^2+(\lambda_2 a_2)^2}} \\
m_3 &= 0.
\end{align*}
Therefore, the function $f$, which achieves this minimal value, is $n_m f_m +n_l f_l$
where 
$$n_m=\frac{-\lambda_4\rho_m(\PXa;\W)b_1}{\sqrt{\left(\lambda_2 a_1+\lambda_4\rho_m(\PXa;\W)b_1\right)^2+(\lambda_2 a_2)^2}}$$
and
$$n_l=\frac{-\lambda_2}{\sqrt{\left(\lambda_2 a_1+\lambda_4\rho_m(\PXa;\W)b_1\right)^2+(\lambda_2 a_2)^2}}.$$
\end{proof}

\subsection{Proof of Theorem~\ref{thm::SimplifyBasis}}
\label{Appendix::ProofBasis}

\begin{proof}
Suppose that $g_m(0)=a\geq 0$ and $g_m(1)=b$. Then $\EE{g_m(Y)|S=0}=a(1-p)+bp=0$ which implies that $b=\frac{-(1-p)}{p}a$. Since $\EE{g_m(Y)^2|S=0}=a^2(1-p)+b^2p=a^2(1-p)+a^2\frac{(1-p)^2}{p}=1$, then $a=\sqrt{\frac{p}{1-p}}$.

Next,
\begin{equation*}
\begin{aligned}
&f_m(\bx)\\
&=\frac{\EE{g_m(Y)|X=\bx,S=0}}{\rho_m(\PXa;\W)}\\
&=\frac{g_m(1)\W(1|\bx)+g_m(0)\W(0|\bx)}{\rho_m(\PXa;\W)}\\
&=\frac{g_m(1)-g_m(0)}{\rho_m(\PXa;\W)}\W(1|\bx)+\frac{g_m(0)}{\rho_m(\PXa;\W)}.
\end{aligned}
\end{equation*}

Note that 
\begin{equation*}
\begin{aligned}
\log\frac{\PXa(\bx)}{\PXb(\bx)}
&= \log\frac{P_{S|X}(0|\bx)P_S(1)}{P_{S|X}(1|\bx)P_S(0)}\\
&= \log\frac{P_{S|X}(0|\bx)}{P_{S|X}(1|\bx)}+\log\frac{P_S(1)}{P_S(0)}.
\end{aligned}
\end{equation*}

\begin{equation*}
\begin{aligned}
f_l(\bx)
&=\log\frac{\PXa(\bx)}{\PXb(\bx)} - \EE{\log\frac{\PXa(X)}{\PXb(X)}\Bigg|S=0}\\
&=\log\frac{P_{S|X}(0|\bx)}{P_{S|X}(1|\bx)} - \EE{\log\frac{P_{S|X}(0|X)}{P_{S|X}(1|X)}\Bigg|S=0}\\
&=\log\frac{1-P_{S|X}(1|\bx)}{P_{S|X}(1|\bx)} - \EE{\log\frac{1-P_{S|X}(1|X)}{P_{S|X}(1|X)}\Bigg|S=0}.
\end{aligned}
\end{equation*}
\end{proof}

\subsection{Details for Example~\ref{ex::Logistic2}}
\label{Appendix::example_logistic}

When $P_{S|X}(1|\bx)=(1+\exp(\theta_0+\dotprod{\bm{\theta}}{\bx})^{-1}$, then
\begin{equation*}
\begin{aligned}
f_l(\bx)
&=\log\frac{1-P_{S|X}(1|\bx)}{P_{S|X}(1|\bx)} - \EE{\log\frac{1-P_{S|X}(1|X)}{P_{S|X}(1|X)}\Bigg|S=0}\\
&= \theta_0+\dotprod{\bm{\theta}}{\bx} - \EE{\theta_0+\dotprod{\bm{\theta}}{X}\Big|S=0}\\
&= \dotprod{\bm{\theta}}{\bx} - \sum_{i=1}^d \EE{\theta_i X_i\Big|S=0}.
\end{aligned}
\end{equation*}
Next,
\begin{align*}
&\EE{f(X)\log\frac{\PXa(X)}{\PXb(X)}\Bigg|S=0} \\
&=\EE{f(X)\log\frac{P_{S|X}(0|X)}{P_{S|X}(1|X)}\Bigg|S=0}\\
&\quad + \EE{f(X)\log\frac{P_S(1)}{P_S(0)}\Bigg|S=0}\\
&= \EE{f(X)\left(\theta_0+ \dotprod{\bm{\theta}}{X}\right)|S=0}\\
&= \theta_0\EE{f(X)|S=0}+\sum_{i=1}^d \theta_i\EE{f(X)X_i|S=0}\\
&= \sum_{i=1}^d \theta_i\EE{f(X)X_i|S=0}.
\end{align*}
Similarly, if we also assume that $P_{S|Y}(1|y)=(1+\exp(\gamma_0+\gamma_1y))^{-1}$, then
\begin{align*}
\EE{g(Y)\log\frac{\PYa(Y)}{\PYb(Y)}\Bigg|S=0} &= \gamma_1\EE{g(Y)Y|S=0}.
\end{align*}
Thus, we can express,
\begin{align*}
&\Delta_{\blambda}\left(f\right) \\
&=\lambda_2\sum_{i=1}^d\theta_i\EE{f(X)X_i|S=0} +\lambda_4\gamma_1\EE{g(Y)Y|S=0}.
\end{align*}

\end{document}